\documentclass[twoside,twocolumn,9pt]{article}
\usepackage{extsizes}
\usepackage[super,sort&compress,comma]{natbib} 
\usepackage[version=3]{mhchem}
\usepackage[left=1.5cm, right=1.5cm, top=1.785cm, bottom=2.0cm]{geometry}
\usepackage{balance}
\usepackage{widetext}
\usepackage{times,mathptmx}
\usepackage{sectsty}
\usepackage{graphicx} 
\usepackage{lastpage}
\usepackage[format=plain,justification=raggedright,singlelinecheck=false,font={stretch=1.125,small,sf},labelfont=bf,labelsep=space]{caption}
\usepackage{float}
\usepackage{fancyhdr}
\usepackage{fnpos}
\usepackage[english]{babel}
\usepackage{lipsum}
\usepackage{array}
\usepackage{droidsans}
\usepackage{charter}
\usepackage[T1]{fontenc}
\usepackage[usenames,dvipsnames]{xcolor}
\usepackage{setspace}
\usepackage[compact]{titlesec}

\usepackage{subfigure}

\usepackage{epstopdf}

\definecolor{cream}{RGB}{222,217,201}

\begin{document}

\pagestyle{fancy}
\thispagestyle{plain}
\fancypagestyle{plain}{

\fancyhead[C]{\includegraphics[width=18.5cm]{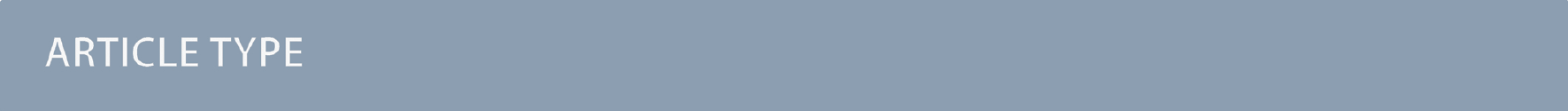}}
\fancyhead[L]{\hspace{0cm}\vspace{1.5cm}\includegraphics[height=30pt]{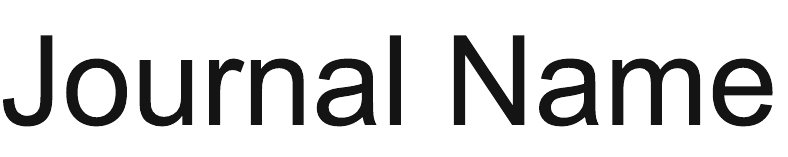}}
\fancyhead[R]{\hspace{0cm}\vspace{1.7cm}\includegraphics[height=55pt]{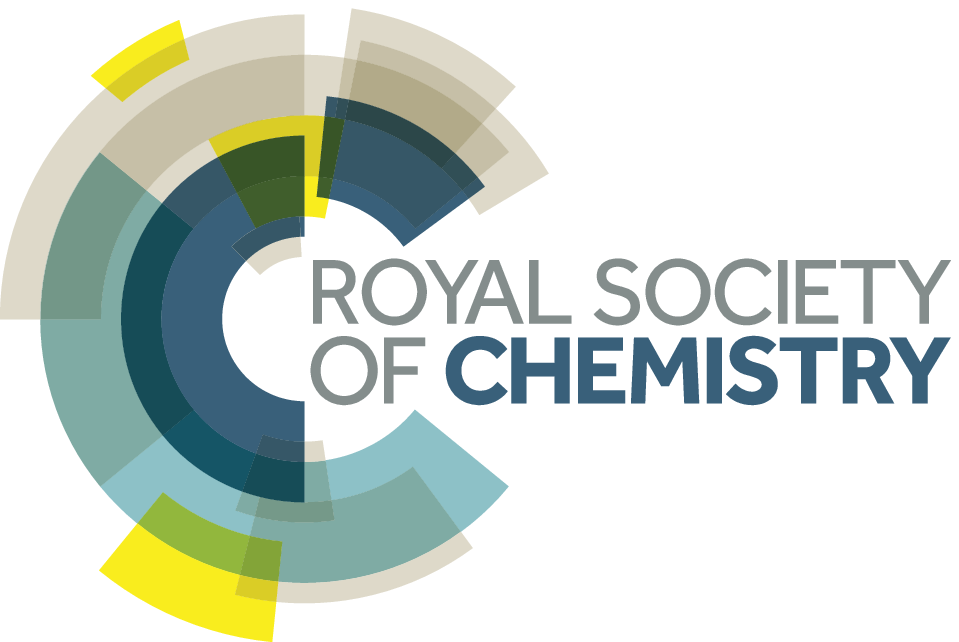}}
\renewcommand{\headrulewidth}{0pt}
}

\makeFNbottom
\makeatletter
\renewcommand\LARGE{\@setfontsize\LARGE{15pt}{17}}
\renewcommand\Large{\@setfontsize\Large{12pt}{14}}
\renewcommand\large{\@setfontsize\large{10pt}{12}}
\renewcommand\footnotesize{\@setfontsize\footnotesize{7pt}{10}}
\renewcommand\scriptsize{\@setfontsize\scriptsize{7pt}{7}}
\makeatother

\renewcommand{\thefootnote}{\fnsymbol{footnote}}
\renewcommand\footnoterule{\vspace*{1pt}%
\color{cream}\hrule width 3.5in height 0.4pt \color{black} \vspace*{5pt}} 
\setcounter{secnumdepth}{5}

\makeatletter 
\renewcommand\@biblabel[1]{#1}            
\renewcommand\@makefntext[1]%
{\noindent\makebox[0pt][r]{\@thefnmark\,}#1}
\makeatother 
\renewcommand{\figurename}{\small{Fig.}~}
\sectionfont{\sffamily\Large}
\subsectionfont{\normalsize}
\subsubsectionfont{\bf}
\setstretch{1.125} 
\setlength{\skip\footins}{0.8cm}
\setlength{\footnotesep}{0.25cm}
\setlength{\jot}{10pt}
\titlespacing*{\section}{0pt}{4pt}{4pt}
\titlespacing*{\subsection}{0pt}{15pt}{1pt}

\fancyfoot{}
\fancyfoot[LO,RE]{\vspace{-7.1pt}\includegraphics[height=9pt]{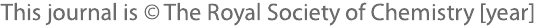}}
\fancyfoot[CO]{\vspace{-7.1pt}\hspace{13.2cm}\includegraphics{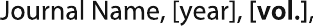}}
\fancyfoot[CE]{\vspace{-7.2pt}\hspace{-14.2cm}\includegraphics{head_foot/RF}}
\fancyfoot[RO]{\footnotesize{\sffamily{1--\pageref{LastPage} ~\textbar  \hspace{2pt}\thepage}}}
\fancyfoot[LE]{\footnotesize{\sffamily{\thepage~\textbar\hspace{3.45cm} 1--\pageref{LastPage}}}}
\fancyhead{}
\renewcommand{\headrulewidth}{0pt} 
\renewcommand{\footrulewidth}{0pt}
\setlength{\arrayrulewidth}{1pt}
\setlength{\columnsep}{6.5mm}
\setlength\bibsep{1pt}

\makeatletter 
\newlength{\figrulesep} 
\setlength{\figrulesep}{0.5\textfloatsep} 

\newcommand{\topfigrule}{\vspace*{-1pt}%
\noindent{\color{cream}\rule[-\figrulesep]{\columnwidth}{1.5pt}} }

\newcommand{\botfigrule}{\vspace*{-2pt}%
\noindent{\color{cream}\rule[\figrulesep]{\columnwidth}{1.5pt}} }

\newcommand{\dblfigrule}{\vspace*{-1pt}%
\noindent{\color{cream}\rule[-\figrulesep]{\textwidth}{1.5pt}} }

\makeatother

\twocolumn[
  \begin{@twocolumnfalse}
\vspace{3cm}
\sffamily
\begin{tabular}{m{4.5cm} p{13.5cm} }

\includegraphics{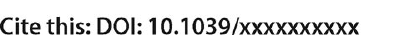} & \noindent\LARGE{\textbf{Role of structural fluctuations and environmental noise in the electron/hole separation kinetics at organic polymer bulk-heterojunction interfaces.
}} \\
\vspace{0.3cm} & \vspace{0.3cm} \\

& \noindent\large{Eric R. Bittner $^{\ast}$\textit{$^{a}$} and Allen Kelley \textit{$^{a}$}  } 
\\
\includegraphics{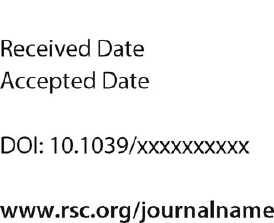} & \noindent\normalsize{
We investigate the   electronic  dynamics of model organic photovoltaic (OPV) system 
consisting of polyphenylene vinylene (PPV) oligomers and [6,6]-phenyl 
C61-butyric acid methylester (PCBM) blend using a 
mixed molecular mechanics/quantum mechanics (MM/QM) approach.  Using a heuristic model that connects 
energy gap fluctuations to the average electronic couplings and 
decoherence times, we provide an estimate of the 
state-to-state internal conversion rates within the manifold of the lowest few electronic excitations.
We find that the lowest few excited states of a model interface are rapidly mixed by C=C bond fluctuations 
such that the system can sample both intermolecular charge-transfer 
and charge-separated electronic configurations on a time scale 
of 20fs.   Our simulations support an emerging picture of carrier 
generation in OPV systems in which interfacial 
electronic states can rapidly decay into charge-separated and current
producing states via coupling to vibronic 
degrees of freedom.
} 

\end{tabular}

 \end{@twocolumnfalse} \vspace{0.6cm}

  ]

\renewcommand*\rmdefault{bch}\normalfont\upshape
\rmfamily
\section*{}
\vspace{-1cm}


\footnotetext{\textit{$^{a}$~Department of Chemistry and Centre for Quantum Engineering, University of Houston, Houston, TX 77204, USA Fax: 713-743-2709; Tel: 832-842-8849; E-mail: bittner@uh.edu}}

\footnotetext{$\ast$ Corresponding author.}

\rmfamily 

\section{Introduction.}  
The power conversion efficiencies of highly optimized 
organic polymer-based photovoltaic (OPV) cells exceeds
10\% under standard solar illumination~\cite{He:2012uq} with reports of 
efficiencies as high as 12\% in multi-junction OPVs. 
This boost in efficiency indicates that mobile charge carriers can be generated 
efficiently in well-optimized organic heterostructures;
however, the underlying mechanism for converting highly-bound Frenkel excitons into mobile 
and asymptotically free photocarriers 
remains elusive in spite of vigorous, multidisciplinary research activity. \cite{Tong2010,doi:10.1021/jz4010569,Gelinas:2013fk,Collini16012009,doi:10.1021/jp4071086,PhysRevLett.114.247003,Provencher:2014aa,Engel:2007aa,Beljonne:2005,tamura:021103,Grancini:2013uq,C3FD20142B,tamura:107402,Sheng2012,yang:045203,ja4093874,doi:10.1021/jp104111h} 
Much of the difficulty in developing a comprehensive picture stems from our lack of detailed understanding 
of the electronic energy states at the interface between donor and 
acceptor materials  and how these states are 
influenced by small--but significant--fluctuations in the molecular structure.

\begin{figure}[h]
\includegraphics[width=\columnwidth]{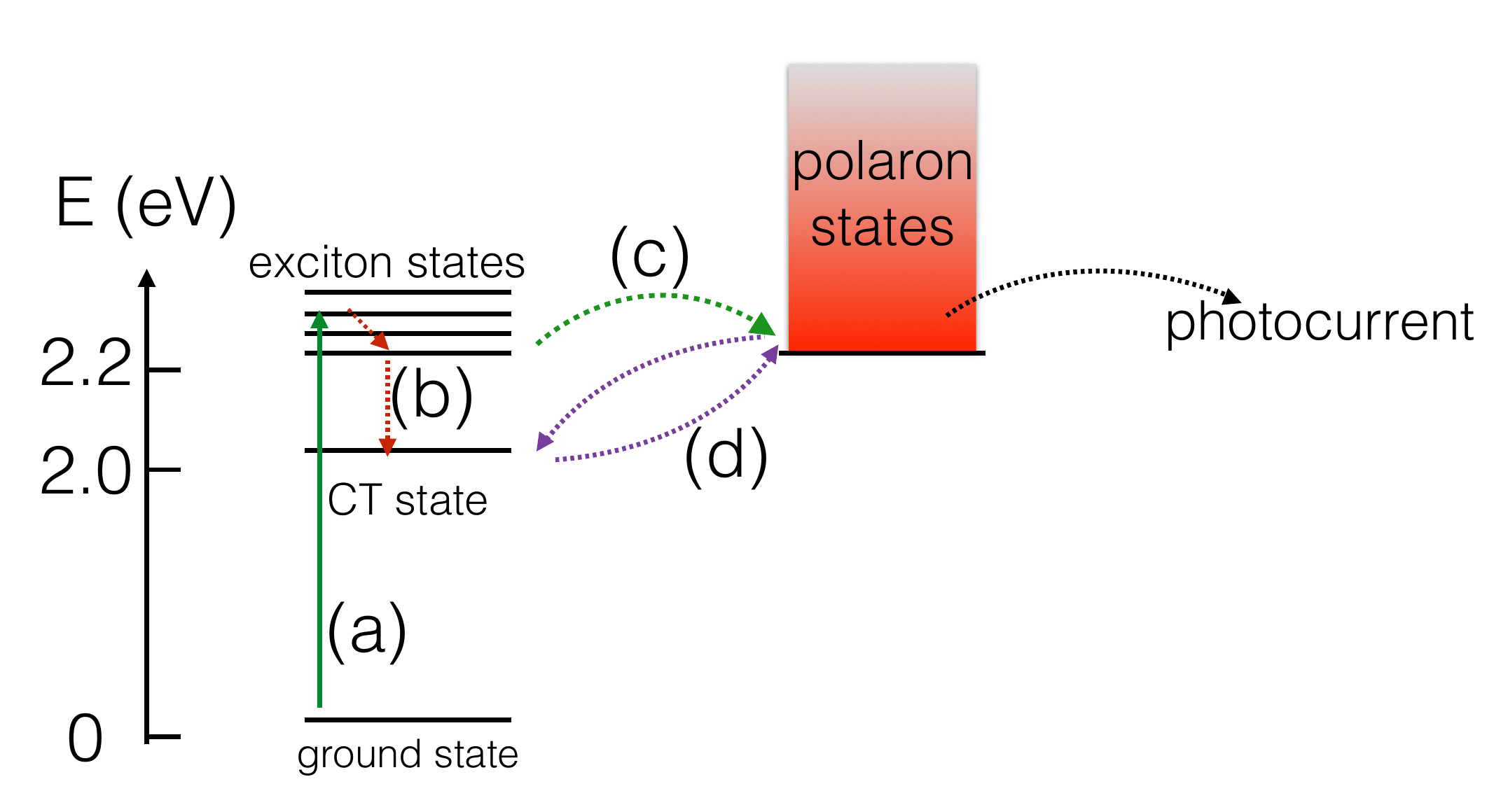}
\caption{Jablonski diagram  showing relative energetic positions of excited states in a bulk-heterojunction system.  The discrete states on the left are assumed to be localised 
excitons or charge-transfer states while the polaron states are assumed to be a quasi-continuum of
mobile charge-separated states within the bulk. The latter of these  are presumed to be responsible for producing photocurrent from a given device.  }
\label{jablonski}
\end{figure}

The generic photophysical pathways that underlie the generation 
of mobile charge carriers in an OPV solar cell
are sketched in Fig.~\ref{jablonski}.   The absorption of a photon by the 
material produces a $\pi-\pi^{*}$ excitation (exciton) 
within the bulk (a) that can migrate and diffuse via 
F{\"o}rster energy transfer processes. Once in close proximity to a bulk-heterojunction 
interface, energetic off-sets between the respective HOMO and LUMO 
levels of adjacent donor and acceptor molecules provide the 
necessary driving force to separate an exciton into a localized charge transfer (CT) state (b) which typically lies 0.25 to 0.4 eV lower in energy. 
Alternatively, an exciton may dissociate directly 
via tunneling into charge-separated (CS) or polaron states (c) which may subsequently evolve 
to contribute to the photocurrent or undergo geminate or non-geminate 
recombination to form CT states (d).   
We distinguish CT states from CS states by 
whether or not the donor and acceptor species are in direct contact (CT) or separated by one or 
more intermediate molecules (CS).   

Ultrafast spectroscopic measurements on OPV systems have reported that charged photoexcitations 
are  generated on $\leq 100$-fs timescales
~\cite{Sariciftci:1994kx,Banerji:2010vn,Tong2010,Sheng2012,Jailaubekov:2013fk,Grancini:2013uq,Banerji:2013ej}; 
however, full charge separation to produce free photocarriers is expected to be energetically expensive given the strong Coulombic 
attraction between electrons and holes due to the low dielectric constant in molecular semiconductors.   
Nonetheless,  experiments  by G\'elinas {\em et~al.}, in which Stark-effect signatures in transient absorption spectra 
were analysed to probe the local electric field as charge separation proceeds, indicate that
 electrons and holes separate by as much as  $40$\AA\ over the first 100\,fs\ and 
 evolve further on picosecond timescales to produce unbound and hence freely mobile charge pairs~\cite{Gelinas:2013fk}. 
 Concurrently, transient resonance-Raman measurements by Provencher {\em et~al.}  demonstrate clear polaronic 
 vibrational signatures on sub-100-fs on the polymer backbone, with very limited molecular reorganization
  or vibrational relaxation following the ultrafast step\cite{Provencher:2014aa}.
 Such rapid through-space charge transfer 
  between excitons on the polymer backbone and acceptors across the heterojunction would 
  be difficult to rationalize within Marcus theory using a localised basis without invoking unphysical distance 
  dependence of tunnelling rate constants~\cite{Barbara:1996uc} and appear to be a
 common feature of organic polymer bulk heterojunction systems.
 
 \begin{figure}
\includegraphics[width=0.75\columnwidth]{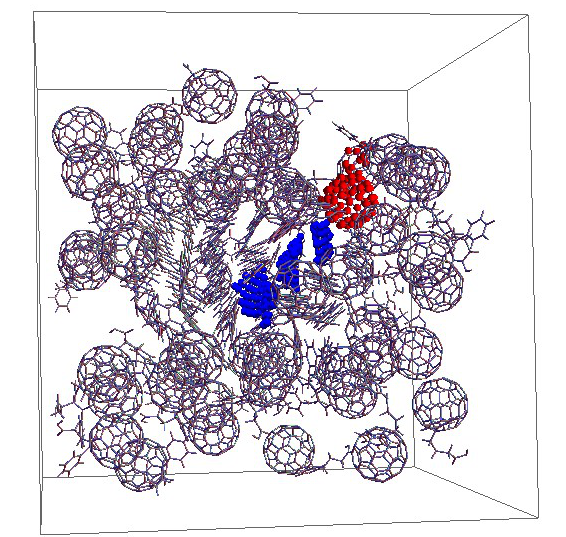}
\caption{Snapshot of the MD simulation cell  containing 50 PCMB molecules and 25 PPV oligomers 
following equilibration at 100 K and 1bar of pressure.  The red and blue highlighted molecules denote the $\pi$-active units
in our system.} 
\label{fig:1}
\end{figure}

 Polymer microstructural probes have 
revealed general relationships between disorder, aggregation and electronic properties in polymeric semiconductors~\cite{Noriega:2013uq}.   
Moreover, aggregation (ordering) can be perturbed by 
varying the blend-ratio and composition of donor and acceptor polymers~\cite{doi:10.1021/jp104111h}.
 On one hand, energetic disorder at the interface would provide a
free-energy gradient for localised charge-transfer states to escape to the 
asymptotic regions.  In essence, the localised polarons in the interfacial region
could escape into bands of highly mobile polarons away from the heterojunction region~\cite{doi:10.1021/nl302172w}.  On the other hand, 
energetic disorder in the regions away from the interface would provide an entropic 
driving force by increasing the density of localised polaron states
away from the interfacial region, allowing the polarons to hop or diffuse away from the 
interface before recombination could take place~\cite{doi:10.1021/jp2083133}.
 It has also been suggested that in polymer/fullerene blends, interfacial exciton fission is facilitated by 
charge-delocalisation along the interface which provide a lower barrier for fission with the excess
energy provided by thermally-hot vibronic dynamics\cite{ja4093874}.
Finally, a report by Bakulin~{\em et~al.} indicates that when relaxed charge-transfer-excitons are pushed with an infrared pulse,
 they increase photocurrent via delocalised states rather than by energy gradient-driven 
hopping~\cite{Bakulin16032012}. 

Recent MD simulations by Fu~{\em et~al.} of a model Squarene/Fullerene OPV cell indicate that the degree of 
disorder at the interface directly affects couplings and hence golden-rule transition rates between the ground and excited states. 
The disorder in the system is at least partly introduced by thermal annealing and the interactions between the Squaraine and Fullerene layers. 
Their simulations indicate that even at 300K, the thermal motions of the molecules at the interface 
can be quite profound and the degree of disorder inherent around the interface can greatly affect the formation of 
electron/hole pairs.\cite{Fu:2014fu}  
 
Finally, we recently presented a fully quantum dynamical model of photo-induced
charge-fission at a polymeric type-II heterojunction interface. \cite{Bittner:2014aa} Our model supposes that the energy level 
fluctuations due to bulk atomic motions create resonant conditions for coherent separation of 
electrons and holes via long-range tunnelling.  Simulations based upon lattice models reveal that such fluctuations
lead to strong quantum mechanical coupling between excitonic states produced near the interface 
and unbound electron/hole scattering states resulting in a strong enhancement of the decay rate of photo excitations into 
unbound polaronic states.

In this work, we employ the our model in conjunction with an atomistic quantum/molecular dynamics 
study of a model donor-acceptor blend to provide an estimate of the interstate transition rates with the excited-state manifold.
Our simulations combine a molecular dynamics description of the atomic motions coupled to an {\em on the fly} description of the 
excited state $\pi$-electronic structure.   By analysing the energy gap fluctuations between excited states, we provide a 
robust estimate of both the interstate electronic couplings,  decoherence times, and transition rates.  We begin with a
brief overview of our model and then describe our results.

\subsection{Heuristic model}
A simple model
for considering the role of the environment can 
be developed as follows\cite{Haken:73,Bittner:2014aa}.   
Consider a two-state quantum system with coupling $\lambda$ in which the 
energy gap $\Delta(t)$ fluctuates in time about its average $\Delta_{o}$  and $\langle\Delta(t)\Delta(0)\rangle = \sigma_{\Delta}^{2}$.
In a two-state basis of $\{|0\rangle, |1\rangle \} $ the Hamiltonian for this can be written as 
\begin{eqnarray}
H = \frac{\Delta(t)}{2}\hat \sigma_z + \lambda \hat\sigma_x
\label{ham1}
\end{eqnarray}
where $\hat\sigma_k$ are Pauli matrices.  Note that Eq. ~\ref{ham1}  can be transformed such that 
fluctuations are in the off-diagonal coupling 
\begin{eqnarray}
H = \frac{\Delta_o}{2}\hat \sigma_z + \delta V(t) \hat\sigma_x
\end{eqnarray}
where $\Delta_o = \overline\Delta + \lambda$  and $\overline{\delta V(t)}= 0$. 
The fluctuations in the electronic energy levels are attributed to thermal and bond-vibrational motions of polymer chains
and can be related to the spectral density, $S(\omega)$ via
\begin{eqnarray} 
 \bar V^{2} = \overline{\delta V^{2}(t)}  = \int_{-\infty}^{+\infty}\frac{d\omega}{2\pi} S(\omega).
 \label{variance}
\end{eqnarray} 
Averaging over the environmental noise, we can write the average energy gap as 
 $\hbar\bar\Omega =\sqrt{\Delta_o^{2} + \bar V^{2}}$
 with eigenstates
 \begin{eqnarray}
|\psi_+\rangle &=& \cos\theta | 0\rangle   +  \sin\theta |1 \rangle \\
|\psi_-\rangle &=& -\sin\theta | 0\rangle   +  \cos\theta |1 \rangle
\end{eqnarray}
where  $\tan2\theta = |\bar V|/\Delta_o$ defines  mixing angle between original kets.
Consequently, by analysing energy gap fluctuations, we can obtain an 
estimate of both the coupling between states as well as transition rates.

\begin{figure}
\subfigure[]{\includegraphics[width=0.45\columnwidth]{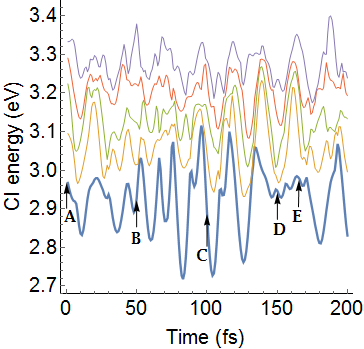}}
\subfigure[]{\includegraphics[width=0.45\columnwidth]{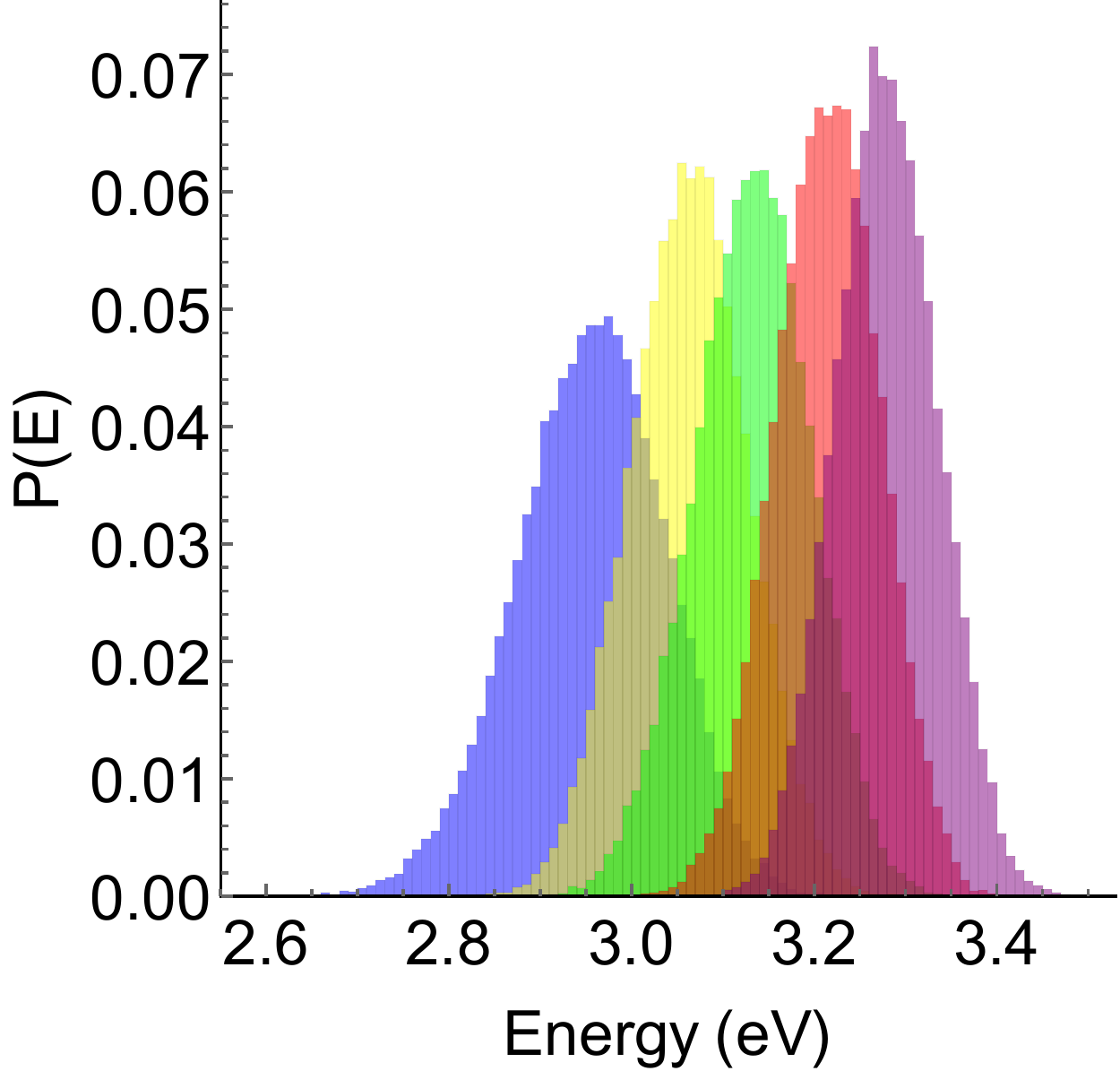}}\\
\subfigure[]{\includegraphics[width=0.75\columnwidth]{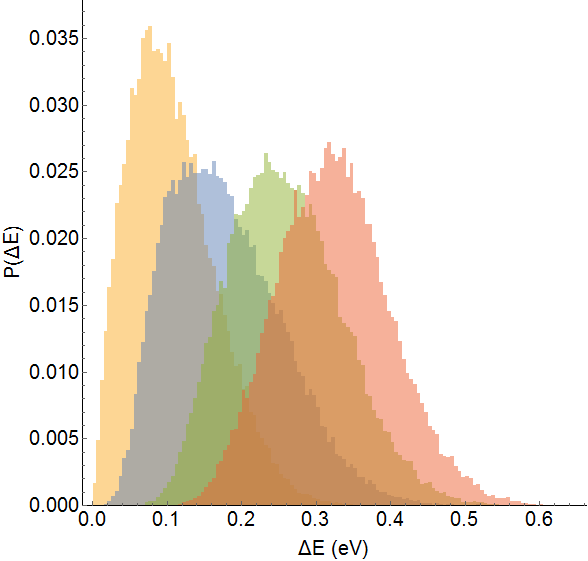}} 
\caption{(a)Single CI (SCI) energies for the first 50 fs following excitation to the lowest SCI state.  (b)  Histogram distribution of the 5 lowest excitation energy
 levels over a 50ps simulation.
Throughout the simulation, the lowest lying state (in blue) remained populated and varied in character from 
excitonic to charge-separated. 
Labels A-E refer to the configurations shown in Fig. 5.
 The color of each component corresponds to specific SCI eigenvalue depicted in (a).
 (c) Histogram of energy gaps between the first excited and higher-lying excited states. 
Gold: $1\to2$, Blue: $1\to 3$, Green: $1\to4$,  Red: $1\to 5$.
} \label{fig:ci-energies}
\end{figure}

\subsection{Noise-averaged kinetics}
To estimate the average transition rate between states, 
we first write the equations of motion for the reduced density matrix for a two level 
system coupled to a dissipative environment,
\begin{eqnarray}
\dot \rho_{11} &=& -\frac{i}{\hbar}\bar V (\rho_{21}-\rho_{12}) - \frac{1}{\tau_{1}}\rho_{11}  \nonumber \\
\dot \rho_{22} &=& \frac{i}{i\hbar}\bar V (\rho_{21}-\rho_{12}) - \frac{1}{\tau_{2}}\rho_{22}\nonumber \\
\dot \rho_{12} &=& \frac{i}{\hbar}\bar V (\rho_{22}-\rho_{11}) - \frac{1}{T_{d}}\rho_{12} + \frac{i\Delta_{o}}{\hbar}\rho_{12} 
\label{density} \\
\dot \rho_{21} &=& \frac{i}{\hbar}\bar V (\rho_{22}-\rho_{11}) - \frac{1}{T_{d}}\rho_{21} - \frac{i\Delta_{o}}{\hbar}\rho_{21} \nonumber 
\end{eqnarray}
where we have introduced $\tau_{1}$ and $\tau_{2}$ as the natural lifetimes of each state and $T_{d}$ as the decoherence time
for the quantum superposition, which in turn, can be related to the spectral density via $T_{d}^{-1} = \bar V/\hbar$.
Taking $T_{d}$ to be short compared to the lifetimes of each state, we can write the population of the initial state as
\begin{eqnarray}
\rho_{11}(t) =\exp\left[- \left(\frac{1}{\tau_{1}} - k \right)t \right]
\end{eqnarray}
where $k$ is average state-to-state transition rate.   Integrating this over time, we obtain an equation of the form
\begin{eqnarray}
\int_{0}^{\infty} \rho_{11}(t)dt = \left( \frac{1}{\tau_{1}} + k \right)^{-1}
\end{eqnarray}
suggesting a form for the exact solution of Eqs.\ref{density}.  Taking the Laplace transform of Eqs.\ref{density}
and assuming that our initial population is in state 1 ($\rho_{11}(0) = 1$), Eqs.\ref{density} become a series of 
algebraic equations
\begin{eqnarray}
-1 &=& -\frac{i}{\hbar}\bar V (\tilde \rho_{21}-\tilde\rho_{12}) - \frac{1}{\tau_{1}}\tilde\rho_{11}  \nonumber \\
0 &=& \frac{i}{\hbar}\bar V (\tilde\rho_{21}-\tilde\rho_{12}) - \frac{1}{\tau_{2}}\tilde\rho_{22}\nonumber \\
0 &=& \frac{i}{\hbar}\bar V (\tilde\rho_{22}-\tilde\rho_{11}) - \frac{1}{T_{d}}\tilde\rho_{12} + \frac{i\Delta_{o}}{\hbar}\tilde\rho_{12} \\
0 &=& -\frac{i}{\hbar}\bar V (\tilde\rho_{22}-\tilde\rho_{11}) - \frac{1}{T_{d}}\tilde\rho_{21} - \frac{i\Delta_{o}}{\hbar}\tilde\rho_{21} \nonumber 
\label{laplacedensity}
\end{eqnarray}
which after a bit of  algebra  gives a rate constant of the form
 \begin{eqnarray}
k &=& 2\frac{\bar V^{2}}{\hbar^{2}} \frac{T_{d}}{(T_{d}\Delta_{o}/\hbar)^{2} + 1} 
\label{rateEQ}
\end{eqnarray}
The average rate vanishes in the limit of rapid decoherence ($T_{d}\to 0$). 
This is the quantum Zeno effect whereby rapid quantum
measurements on the system by the environment at a rate given by $T_{d}^{-1}$  collapses
the quantum  superposition formed due to the interactions and thereby
suppresses transitions between states.
On the other hand,  transient coherences can
facilitate quantum transitions between otherwise weakly coupled states. 

For the sake of connecting this to the photophysical dynamics of a OPV heterojunction,
let us assume that one state corresponds to a CT state and the other corresponds to a CS state. 
When the fluctuations are weak, $\bar V \ll \Delta_{o}$,  the original kets $|CT\rangle$ and $|CS\rangle$ provide a good zeroth description of actual 
eigenstates of the system and the coupling can be treated as a weak perturbation.    
However, when the off-diagonal couplings are comparable to the average 
gap,  even  low-lying CT states may be brought into strong coupling with the CS states.  

The implication of this heuristic model is that one can obtain the input to the rate 
equation (Eq. ~\ref{rateEQ}) from mixed quantum classical simulations that 
take into account the excited state populations. 
Here we report on such simulations of a model OPV system consisting of a blend of 
fullerene and polyphenylene vinylene oligomers as depicted in Fig.~\ref{fig:1}.  
Our simulation method employs an atomistic description of the 
nuclear dynamics described by a force-field that responds to changes in the local $\pi$ electronic structure of a 
sub-set of molecules with in the simulation cell.  
We restrict the excited state population to the lowest $\pi-\pi^{*}$excitation as to simulate the long-time fate of
a singlet CT state prepared via either photoexcitation or charge recombination. 
By analysing the energy gaps between electronic adjacent states and the character of
the excited states in terms of electron-hole configuations we can deduce how 
small vibronic motions of the polymer chains modulate the electronic coupling and 
induce charge-separation.


\begin{figure*}[t]
\subfigure[0 fs]{\includegraphics[width=0.6\columnwidth]{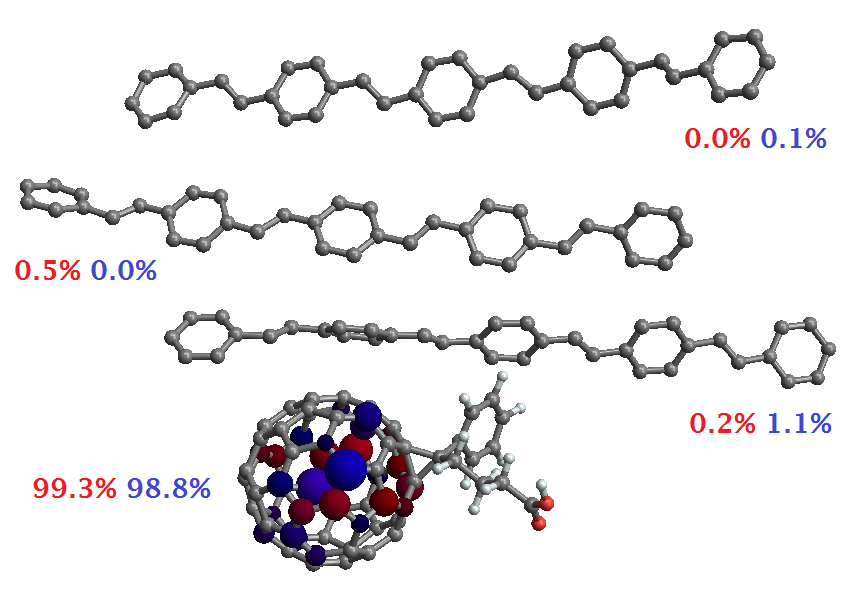}}
\subfigure[50 fs]{\includegraphics[width=0.6\columnwidth]{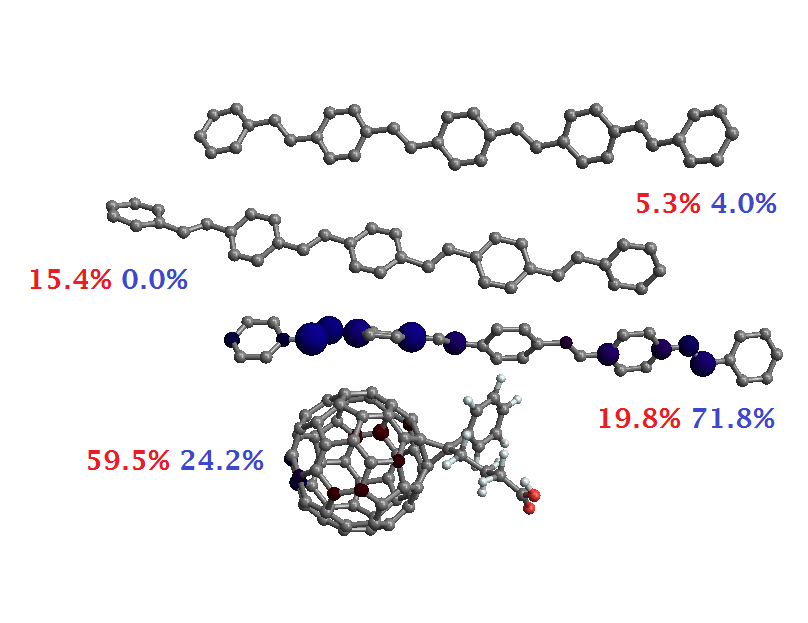}}
\subfigure[100 fs]{\includegraphics[width=0.6\columnwidth]{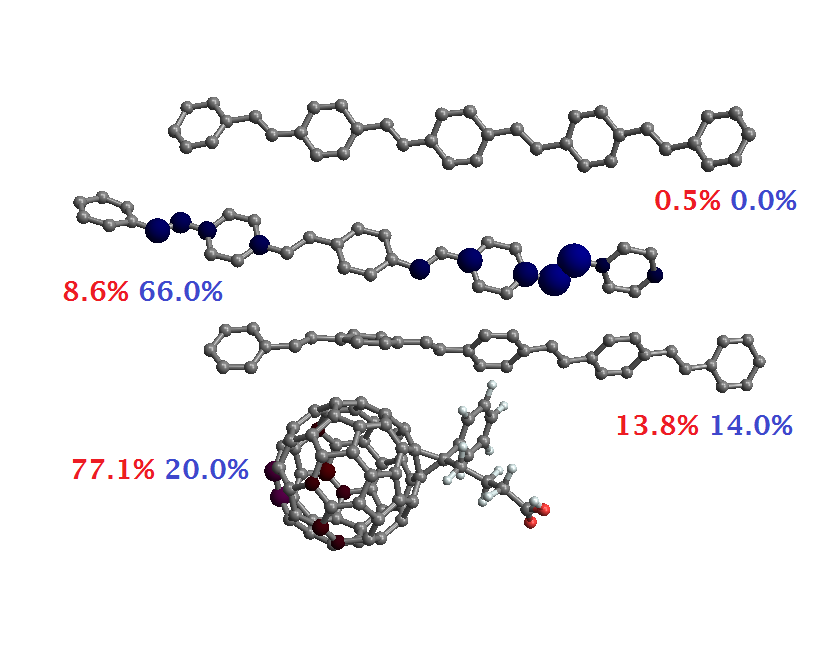}}\\
\subfigure[150 fs]{\includegraphics[width=0.6\columnwidth]{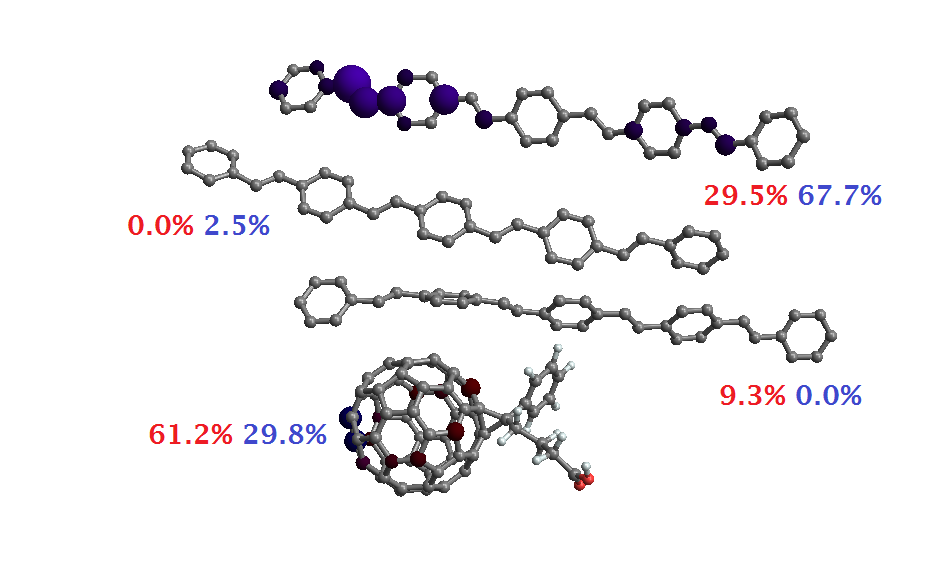}}
\subfigure[100 fs]{\includegraphics[width=0.6\columnwidth]{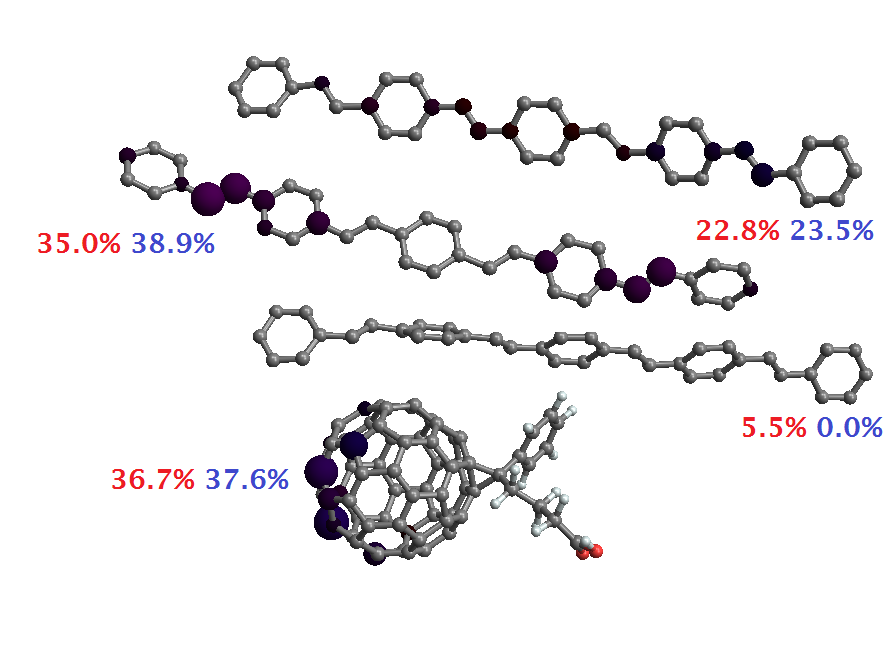}}
\subfigure[150 fs]{\includegraphics[width=0.6\columnwidth]{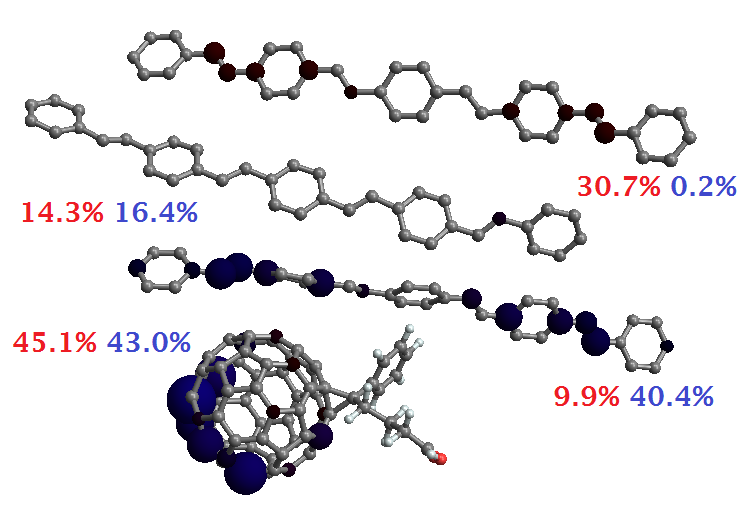}}
\caption{Sequence of states following initial excitation.  In this particular simulation, the initial excited state population was localised on the PCBM.  This 
population rapidly dissociates into charge-transfer and then charge-separated configurations within 100-150 fs.  This scenario appears to be a 
ubiquitous feature of our simulations indicating that fluctuations alone can carry the system from a localized CT state to charge-separated states and eventually 
current-producing polaron states.} 
\label{fig:densplots2}
\end{figure*}

\section{Methods.}   

Our simulations employ a modified version of the TINKER molecular dynamics (MD)  package\cite{ponder:2004}
 in which the MM3\cite{allinger:868} intramolecular bonding parameters were allowed to vary with the local $\pi$-electronic density
 as described by a Parisier-Parr-Pople (PPP) semi-empirical Hamiltonian. 
 \cite{pariser:767,pople:1375}
 Specifically, we assume that the internal bond force constants, bond-lengths, bond angle, bending potentials, and 
 bond torsion parameters are linear functions of the local bond-order.  
 We specifically chose 1 PCBM and 3 nearby PPV oligomers to represent a model bulk-heterojunction in order to study the 
penetration of extended intramolecular electronic states into the bulk region.   
 The remaining molecules in the simulation 
were treated using purely classical force-field. 
 At each step of the simulation,  we compute the Hartree-Fock (HF) ground-state for the $\pi$ system and use configuration interaction (singles)
 (SCI) to describe the lowest few $\pi\to\pi^{*}$ excitations.   Intermolecular
 interactions within the active space were introduced via non-bonding Coulombic  coupling terms and
static dispersion interactions contained within the MM3 forcefield.

Fig.~\ref{fig:1} shows a snapshot of the 
periodic simulation cell  containing 50 PCMB molecules and 25 PPV oligomers 
following equilibration at 100K and 1bar of pressure using classical molecular dynamics (MD)
within the NPT ensemble\footnote{PCBM:Phenyl-C61-butyric acid methyl ester, PPV: Polyphenylene vinylene}.  
The molecules surrounding the 4 $\pi$-active molecules serve as a thermal bath and 
 electronic excitations are confined to the $\pi$-active orbitals.  In total, our $\pi$-active space included  a total of 172  carbon $2p_{z}$ orbitals and we used a total of 10 occupied and 10 unoccupied orbitals to construct electron/hole configurations for the 
 CI calculations.     During the equilibration steps, we assume the system to be in its electronic ground state, 
 after which  we excite the system to the first SCI excited state and allow the system to respond to the change 
 in the electronic density within the adiabatic/Born-Oppenheimer approximation.    It is important to note that the excited state we prepare 
 is not the state which carries the most oscillator strength to the ground
 nor do we account for non-adiabatic surface hopping-type 
transitions in our approach.\cite{tully:1061,tully:562,Granucci:2007}  Our dynamics and simulations
 reflect the longer-time fate of the lowest-lying excited state populations as depicted in step (d) in Fig.~\ref{jablonski}.
The combination of a classical MD forcefield with a semi-empirical description of a select few molecules within the system 
seems to be a suitable compromise between a fully {\em ab initio} approach which would be limited to only a few molecules
and short simulation times and a fully classical MD description which would neglect any 
transient changes in the local electronic density~\cite{Jailaubekov:2013fk}.
In spite of the relative simplicity of our model, the simulations remain quite formidable. 
\section{Results}   

Over the course of a 50 ps simulation,  the lowest lying SCI excitation samples a variety of electronic configurations 
ranging from localised PCBM excitons to charge-separated and charge-transfer states with varying degrees of charge separation. 
In Fig.~\ref{fig:ci-energies}a, we show the lowest few SCI excitation energies following excitation at $t=0$~fs to the lowest SCI state for 
one representative simulation.   The labels refer to snap-shots taken at 50fs intervals to visualize the various 
electronic configurations sampled by our approach in Fig. \ref{fig:densplots2}. 
First, we note that following promotion to the lowest lying SCI at $t=0$ there is very little energetic reorganization or relaxation compared to the 
the overall thermal fluctuations that modulate the SCI eigenvalues.  This can be rationalized since the excitation to the $\pi^{*}$ orbitals at any given time 
is largely delocalised over one or more molecular units and hence the average electron-phonon coupling per C=C bond is quite small.  
The number of avoided crossings that occur between lowest lying states is highly striking, signalling that the internal 
molecular dynamics even at 100K is sufficient to bring these states into regions of strong electronic coupling.  

In Fig.~\ref{fig:ci-energies}b we show a histogram of the energies of the lowest 5 SCI energies accumulated over 50 ps of simulation time following 
promotion to the lowest SCI state.   Focusing upon the lowest lying state (in blue), fluctuations due to molecular dynamics
 can account for nearly 0.25eV of inhomogeneous broadening of this state bringing it (and similarly for the other states) into regions of 
 strong electronic coupling with nearby states which dramatically changes the overall electronic character of the state from purely charge-transfer 
 to purely excitonic on a rapid time-scale. 
 
We next consider the origins of the energy fluctuations evidenced in Fig.~\ref{fig:ci-energies}a.  While we only show a 200fs segment of a much longer 50ps simulation, 
over this period one can see that the CI energies are modulated with a time-period of about 20 fs.  Indeed,  the Fourier transform of the average CI excitation energy
reveals a series of peaks between 1400-1700 cm$^{-1}$ which  corresponds to the C=C stretching modes present in the polymer chains and fullerene 
molecules.  We conclude that  small-scale vibronic fluctuations in the molecular structures and orientations produce significant 
energetic overlap between different electronic configurations
in organic polymer-fullerene heterojunction systems and speculate that this may be the origins of efficient charge
separation in such systems. 

\begin{figure*}[t]
\subfigure[51 fs]{\includegraphics[width=0.6\columnwidth]{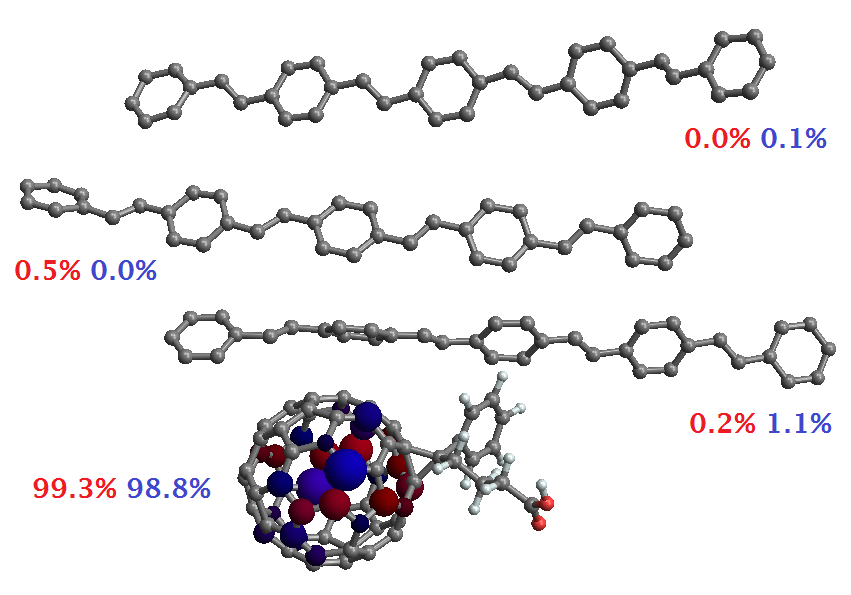}}
\subfigure[52 fs]{\includegraphics[width=0.6\columnwidth]{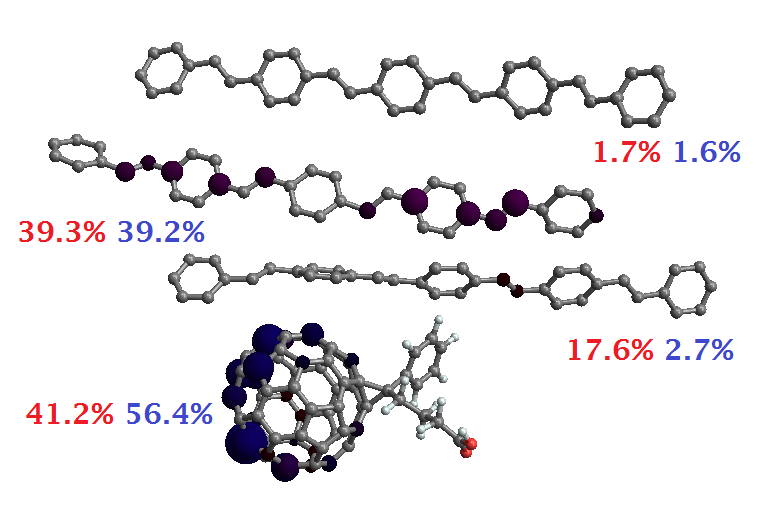}}
\subfigure[53 fs]{\includegraphics[width=0.6\columnwidth]{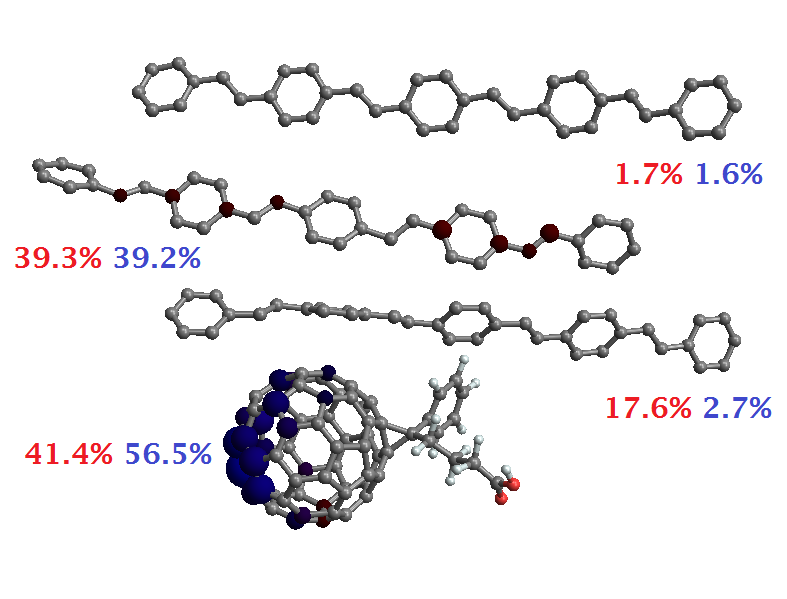}}\\
\subfigure[54 fs]{\includegraphics[width=0.6\columnwidth]{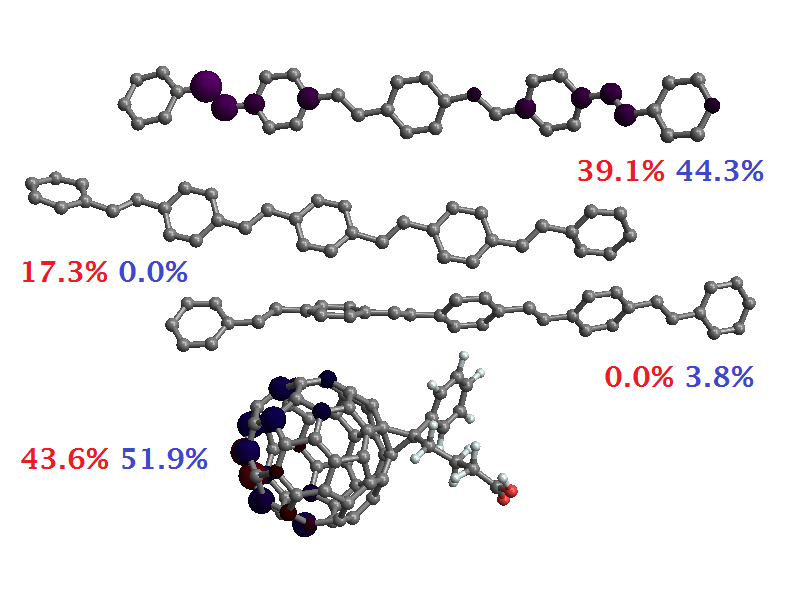}}
\subfigure[55 fs]{\includegraphics[width=0.6\columnwidth]{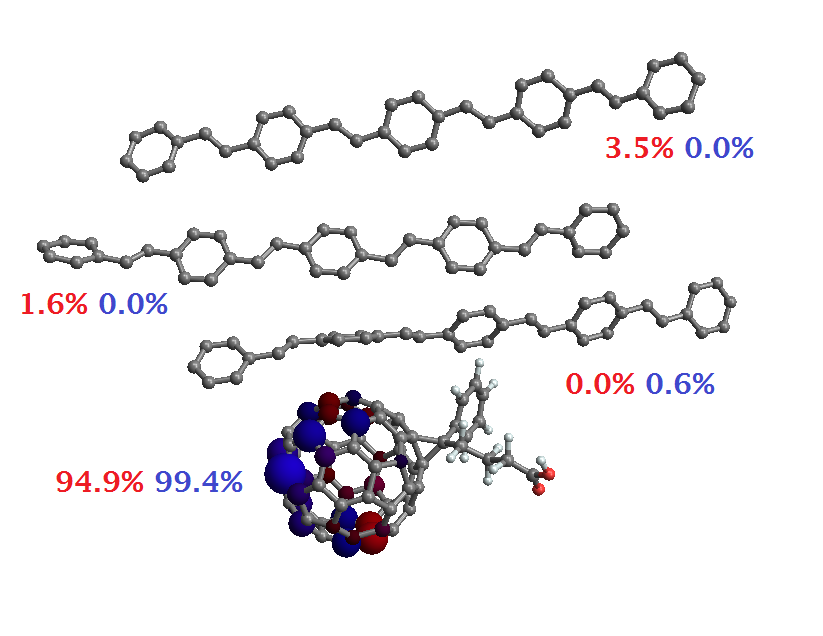}}
\subfigure[56 fs]{\includegraphics[width=0.6\columnwidth]{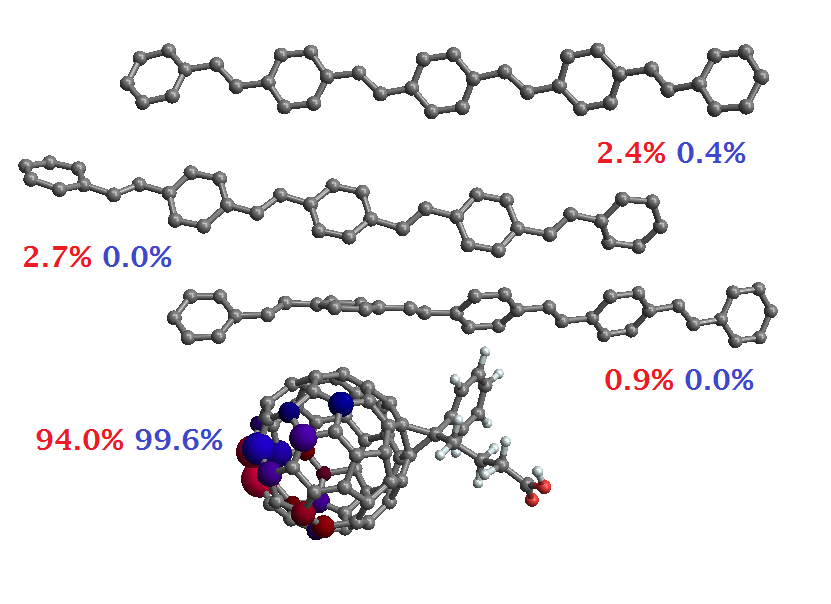}}
\caption{ A 6 fs time history from 51 to 56 fs of the first excited CI state is shown in Fig. ~\ref{fig:ci-energies}. This series resides around an avoided crossing between the first and second excited state. Showing how a CT state at the interface can evolve into exciton states farther into the bulk around one avoided crossing points. After the excited states diverge, the exciton in the bulk collapses back into an exciton localised on the PCBM. } 
\label{fig:densplots1}
\end{figure*}

A variety of states are produced by our model and we categorize them as excitonic (electron/hole on the same molecular unit) occupying
$\approx 50\%$  of the states, charge-transfer (electron on PCMB/hole  on PPV\#1 (nearest to PCBM)) 
occupying $\approx$ 15\% of the states, or charge separated 
(electron on PCBM/hole on PPV \#2 or \#3) occupying $\approx$ 25\% of the states.
These are shown in Fig.~\ref{fig:densplots1} where we indicate the local charge-density of each C2$p_{z}$ orbital by a blue (positive) or red (negative) sphere 
scaled in proportion to the local  charge on each site.

In Fig.~\ref{fig:densplots2}  we show a sequences of snap shots taken every 50 fs following the initial excitation.  
These figures reveal the highly dynamical nature of the lowest lying excitation in which the system samples
excitonic configurations (a), charge-transfer configurations (b,c),  charge-separated states in which the electron and hole are separated by at least one 
polymer chain (d,e), and highly mixed charge-tranfer/excitonic states as in (e,f).   
The scenario depicted here is not an exceptional case within our simulation.  
Here, the initial excited state is 
clearly localized on the PCBM unit with a high degree of excitonic character.  There is clearly some charge separation within the PCBM; however, both the electron and
hole clearly reside on the PCBM.   After 50 fs ({\em c.f.} Fig.~\ref{fig:densplots2}b),  structural fluctuations in part induced by changes
 in electronic density and in part by the thermal fluctuations of the surrounding 
medium bring this state nearly into resonance with the second SCI state inducing charge separation between the PCBM and a nearby PPV oligomer. 
Further fluctuations bring this state into resonance with other charge separated states as shown in Fig.~\ref{fig:densplots2}c, d, and f. 
Because our quantum subspace is restricted to the molecules shown in Fig.\ref{fig:densplots2},  charge separation to more distant regions can not occur.

Fig.~\ref{fig:ci-energies}c shows a histogram of the energy gaps between the first  and higher-lying excited excited
states of our system.  According to our heuristic model, the mean ($\mu$) and variance ($\sigma^{2}$) of the gap distributions can be used as input to estimate the 
state-to-state rate for this system under the mapping that $\mu = \Delta_{o}$ and $\sigma^{2} =\overline{\delta V^{2}} $ as per Eq.~\ref{variance}. 
Furthermore, we take $T_{d}^{-1} \approx\bar V/\hbar$ as an estimate of the decoherence time. The results are shown in Table.\ref{table1}.
The estimated transition rates are consistent with the observation that the system rapidly samples 
a wide number possible configurations over the course of the molecular dynamics simulation.  On average, 
the state-to-state coupling of 70 meV is comparable to the 
average energy gaps between the lowest states.  This brings the system into the regime of strong electronic
coupling.  However, larger  couplings also imply {\em shorter} electronic
decoherence times--which dramatically limit the the ability of charges to separate by 
tunnelling.  In this case, any superposition CI states resulting from the quantum mechanical 
time-evolution of the system would collapse to single eigenstate on a sub-10 fs time-scale
due to the interaction with the vibronic degrees of freedom.   Based upon our model and simulations, 
we estimate that within 20 fs, population in any low-lying electronic state is effectively mixed with 
nearby states simply due to the underlying vibrations of the lattice. 

Fig.~\ref{fig:densplots1} illustrates how the electronic nature of the lowest lying excitation changes
over a short time period.   Shown here are the excess charge-densities at each atomic site 
taken at 1~fs intervals.  At each 1~fs time-step, the electronic character changes wildly 
due to strong coupling between the electronic and vibrational degrees of freedom.

\begin{table}[h]
\caption{Estimated interstate transition rates from Eq.~\ref{rateEQ} and vibronic couplings. }
\begin{center}
\begin{tabular}{c|c|c|c|c|}
Transition   &   $\Delta_{o} $  (eV)  &    $\bar V$  (eV) & $T_{d}$ (fs) &$\bar k^{-1}$ (fs) \\
\hline
$1\to 2$      &   0.11  &  0.06     & 11.41  & 21.8 \\
$1\to 3$      &   0.18  &  0.08     & 8.66   & 16.6   \\
$1\to 4$      &   0.26  &  0.08     & 8.70   & 16.7  \\
\end{tabular}
\end{center}
\label{table1}
\end{table}

\section{Conclusions}     

We present here the results of hybrid quantum/classical simulations of the excited states of a model  polymer/fullerene heterojunction interface. 
Our results indicate that dynamical fluctuations due to both the response of the system to the initial excitation and to thermal noise 
can rapidly change the character of the lowest lying excited states from purely excitonic to purely charge separated over a time scale 
on the order of 100 fs.   In many cases, an exciton localised on the fullerene will dissociate into a charge-transfer state with the hole (or electron) 
delocalised over multiple polymer units before localizing to form a charge-separated state.  While the finite size of our system prevents further 
dissociation of the charges, the results are clearly suggestive that such interstate crossing events driven by bond-fluctuations
can efficiently separate charges to a distance to where their Coulombic attraction is comparable to the thermal energy. 

The nuclear motions most strongly coupled to the electronic 
degrees of freedom correspond to C=C bond stretching modes implying that small 
changes in the local lengths have a dramatic role in modulating the 
electronic couplings between excited states.  Since $kT \ll \hbar\omega$ for these
modes, they should be treated quantum mechanically rather than as classical motions. 
Generally speaking, including quantum zero-point effects into calculation of Fermi golden-rule rates leads to {\em slower} transition rates than those computed using classical correlation functions which implies that the values estimated here are the upper bounds for the actual transition rates.  The results presented here corroborate recent ultrafast 
experimental evidence suggesting that free polarons can form on ultrafast timescales (sub 100fs)  and underscore the dynamical nature of the bulk-heterojunction interface. 
\cite{Banerji:2013ej,ja4093874,Chin:2013fk,Rozzi:2013fk,Noriega:2013uq,Bittner:2014aa,Provencher:2014aa,Gelinas:2013fk,PhysRevLett.114.247003}

\section{ Acknowledgments.} 
The work at the University of Houston was funded in part by the
 National Science Foundation (CHE-1362006) and the Robert A. Welch Foundation (E-1337). 
The authors declare no competing interests.  
We also wish to acknowledge many discussions with Prof. Carlos Silva over the course of this work.

\scriptsize{
\providecommand*{\mcitethebibliography}{\thebibliography}
\csname @ifundefined\endcsname{endmcitethebibliography}
{\let\endmcitethebibliography\endthebibliography}{}

 } 

\end{document}